\documentclass[apj]{emulateapj}
\usepackage{graphicx}
\usepackage{pslatex}
\usepackage{times}
\usepackage{color}
\usepackage{natbib}
\usepackage{hyperref}

\hypersetup{colorlinks=true,citecolor=blue,linkcolor=black,filecolor=black,runcolor=black}

\def\lesssim{\lower.5ex\hbox{$\; \buildrel < \over \sim \;$}}
\def\gtrsim{\lower.5ex\hbox{$\; \buildrel > \over \sim \;$}}

\newcommand{\nH}{\mbox{$n_{\rm H}$}}
\newcommand{\kms}{\mbox{${\rm km\,s^{-1}}$}}
\newcommand{\msun}{\mbox{$\rm M_\odot$}}
  
\newcommand{\mvir}{\mbox{$M_{\rm vir}$}}

\newcommand{\mhalo}{\mbox{$M_{\rm halo}$}}

\newcommand{\cmq}{\mbox{${\rm cm^{-3}}$}}
\newcommand{\zsun}{\mbox{$Z_\odot$}}

\definecolor{gray}{rgb}{0.5,0.5,0.5}

\begin{document}

\title{Formation of globular clusters in atomic-cooling halos via rapid gas condensation and fragmentation 
during the epoch of reionization}

\author{Taysun Kimm$^{1,2}$, Renyue Cen$^1$, Joakim Rosdahl$^3$, \& Sukyoung K. Yi$^4$}
\affil{$^1$ Princeton University Observatory, Peyton Hall, 4 Ivy Lane, Princeton, NJ 08544-1001, USA \\
$^2$ Currently at the Kavli Institute for Cosmology and Institute of Astronomy, Madingley Road, Cambridge CB3 0HA, UK\\
$^3$ Leiden Observatory, Leiden University, P.O. Box 9513, 2300 RA, Leiden, The Netherlands \\
$^4$ Department of Astronomy and Yonsei University Observatory, Yonsei University, Seoul 120-749, Republic of Korea
}

\shorttitle{GCs in a hierarchical universe}
\shortauthors{Kimm, Cen, Rosdahl, \& Yi}
\email{Email: tkimm@ast.cam.ac.uk}

\begin{abstract}
We investigate the formation of metal-poor globular clusters (GCs) at the center of 
two dark matter halos with $\mhalo\sim4\times10^7\,\msun$ at $z>10$ using  
cosmological radiation-hydrodynamics simulations. 
We find that very compact ($\lesssim$ 1 pc) and massive ($\sim6\times10^5\,\msun$) clusters form
rapidly when pristine gas collapses isothermally with the aid of efficient Ly$\alpha$ emission 
during the transition from molecular-cooling halos to atomic-cooling halos. 
Because the local free-fall time of dense star-forming gas is very short ($\ll 1\,{\rm Myr}$),
a large fraction of the collapsed gas is turned into stars before stellar feedback processes
blow out the gas and shut down star formation. 
Although the early stage of star formation is limited to a small region of the central 
star-forming disk, we find that the disk quickly fragments due to metal enrichment from supernovae. 
Sub-clusters formed in the fragmented clouds eventually merge with the main cluster at the center.
The simulated clusters closely resemble the local GCs in mass and size 
but show a metallicity spread that is much wider than found in the local GCs. 
We discuss a role of pre-enrichment by Pop III and II stars as a potential solution to the latter issue. 
Although not without shortcomings, it is encouraging that a naive blind (not tuned) cosmological 
simulation presents a possible channel for the formation of at least some massive GCs.
\end{abstract}

\keywords{galaxies: high-redshift --- galaxies: ISM}

%===============
\section{Introduction}
%===============
Numerical simulations show that a variety of systems can emerge during the 
hierarchical build-up of small structures in an LCDM cosmology. 
Disc-dominated galaxies are formed by the coherent accretion of cold gas in conjunction 
with the removal of low-angular momentum gas via effective stellar feedback  
\citep{pichon11,governato07,dubois14}, while violent relaxation during galaxy 
mergers/interactions leads to the formation of a massive bulge \citep[e.g.,][]{toomre72,naab06}. 
In small halos with $\mvir\lesssim 10^{10}\,\msun$, supernova (SN) explosions seem to
be able to alter the gravitational potential non-adiabatically by blowing out a significant amount of gas, 
producing diffuse dwarf galaxies \citep{pontzen12}. 
However, how dense structures, such as GCs, form in a cosmological context remains
an interesting puzzle.
 
The formation of GCs is usually associated with high-pressure regions 
\citep[$P/k_B \gtrsim 10^{7} {\rm cm^{-3}\, K,}$][]{elmegreen97}, owing to their compact nature.
This implies that  the average density of the GC-forming, isothermal gas cores is very high
($\nH\sim10^{5}-10^{6}\, \cmq$), given the typical temperature of star-forming regions (10-100 K).
Because the free-fall timescale of these clouds ($t_{\rm ff}\sim {\rm 0.1\, Myr}$) is 
an order of magnitude smaller than the typical dispersal timescale due to ionizing radiation
\citep[$\sim {\rm 1-2\, Myr,}$][]{walch12,dale14,sales14} or a SN explosion  
($>3\,{\rm Myr}$), the gas in the high-pressure regions can be efficiently converted into stars, 
although star formation itself is a slow process on galactic scales \citep{kennicutt98}.
An important difference between the GC formation sites and normal star-forming regions is 
that a large amount of gas needs to be accumulated and collapse on a short timescale 
in the former case. If the gas accretion onto the cloud occurs very slowly compared to 
star formation, only a small fraction of the gas would achieve the high pressure and 
it may not form a typical, massive GC with $\sim2\times10^5\,\msun$ 
before stellar feedback suppresses star formation. In this case, subsequent gas 
accretion after the initial burst of star formation is likely to lead to extended star 
formation histories, as in dwarf galaxies \citep[e.g.,][]{tolstoy09}.

Interacting galaxies or gas-rich, clumpy discs in high-$z$ galaxies are thus 
good candidates hosting such high-pressure regions. During a galaxy merger, 
gas is funneled rapidly to the central regions due to gravitational torques, 
triggering starbursts \citep{mihos96}. Tidal arms are also fragmented and 
create gas clumps, providing a favorable condition for cluster formation 
\citep[e.g.,][]{teyssier10,renaud15}. Indeed, young, metal-rich star clusters 
are often observed in nearby interacting galaxies \citep[e.g.,][]{whitmore95,schweizer96,bastian09}.
Massive clumps with $\sim10^8-10^{9}\,\msun$, observed in gas-rich discs 
at high redshifts \citep{genzel08,elmegreen09,swinbank11}, may be forming GCs 
as well \citep{kravtsov05,kruijssen15}. Because these clumps are dense 
($\Sigma_{\rm gas} \sim10-100\,\msun/{\rm pc}^3$) and strongly turbulent 
($\sigma_{\rm gas}\sim10-100\,\kms$), they are likely to host high-pressure sub-clumps 
that can turn into a GC. Based on the assumption that these massive clumps 
share the gas-phase metallicity of their host galaxy, \citet{shapiro10} claims that the 
number of metal-rich GCs observed in intermediate-mass
galaxies with $\sim10^{10}-10^{11}\,\msun$ may be explained by the number of massive clumps,
provided that $\sim$ 700 GCs form per clump and only $\sim$ 2 percent of them survive 
the two-body relaxation-driven tidal evaporation \citep{jordan07}.

On the other hand, the origin of metal-poor GCs has been highly debated.
Some authors argue that metal-poor GCs form essentially through the same processes
as metal-rich GCs in gas-rich, massive galaxies at high redshifts \citep{kravtsov05,kruijssen15},
except that the host galaxy should be less massive to match the lower metallicity of the blue GCs. 
In a similar context, \citet{elmegreen12} proposes that Lyman alpha emitters, which are low-mass, 
dense, and actively star-forming galaxies \citep{pirzkal07,gawiser07,finkelstein07}, 
can produce as many as the massive ($>2\times10^5\,\msun$) blue GCs observed in the local Universe. 
Alternatively, some metal-poor GCs may have formed at the center of 
low-mass halos in the young universe, possibly before reionization ($z>6$) \citep[e.g.,][]{katz13}. 
\citet{peebles84} suggested that metal-poor GCs may arise in extended dark halos 
with $10^8\,\msun$ when a gas cloud of mass $10^6\,\msun$ becomes Jeans unstable 
\citep[see also][]{fall85}.  \citet{bromm02} extended the idea by performing three-dimensional 
hydrodynamic simulations, and speculated that fragmented gas clumps with $10^5\,\msun$ 
in the atomic-cooling halos may turn into GCs. 
\citet{boley09} later showed that two simulated clusters of relatively small 
masses ($\sim 5\times10^4\,\msun$) are indeed formed in a minihalo 
with $5\times10^6\,\msun$ at $z\sim13$ in their cosmological simulations 
with star formation and SN feedback. 
Recently, \citet{trenti15} proposed that a merger of two $10^8\msun$ halos that are pre-enriched 
by SNe from nearby (proto-)galaxies may trigger the formation of a metal-poor GC. 
The main criticism of these dark matter halo (DMH)-based scenarios was that GCs seem to possess no 
extended halos \citep{baumgardt09,conroy11,ibata13}, but it is possible that 
metal-poor GCs form through several different channels as aforementioned. 
Also dynamical modeling of mergers of dark matter halos suggests that tidal stripping can 
remove a substantial amount of dark matter before stellar components are affected \citep[e.g.,][]{mashchenko05,smith15}, 
making this scenario still viable.

The DMH-based scenario is barely studied from a baryonic physics viewpoint 
 in a fully cosmological setup, however \citep[see][for an exception]{boley09}. 
GCs are conventionally known as a simple stellar population (SSP) with homogeneous chemical composition,
although more complex features, such as abundance variations \citep{gratton04,lim15} or multiple stellar 
populations \citep[e.g.,][]{lee99,bedin04,joo13}, are observed. One might speculate that these 
two well-known features (coeval age and chemical homogeneity) 
can easily be reproduced if a gas clump with an appropriate mass and metallicity is turned into a 
cluster instantaneously.
But this is a non-trivial problem in a cosmological context.
First, cosmic accretion supplies fresh gas to a DMH continuously, 
possibly resulting in populations with extended star formation histories rather than an SSP. 
Depending on how rapidly or slowly stellar feedback processes regulate 
gas collapse and star formation, the mass of a star cluster may vary significantly.
Second, if a substantial amount of gas that dominates 
the gravity of the GC-forming regions is removed by stellar feedback before a self-gravitating star cluster 
forms, they are likely to expand non-adiabatically, resulting in a diffuse structure \citep[e.g.,][]{pontzen12}.
An important question is thus whether a massive ($>2\times10^5\,\msun$) and dense structure can ever be 
formed in a cosmological environment where gas accretion appears to occur slowly and continuously.
Third, given that metal enrichment and mixing are unlikely to be homogeneous and instantaneous, 
the interplay between star formation and feedback may produce chemically inhomogeneous
populations of stars. 
The aim of this paper is to examine these three aspects of GC formation by carrying out 
a high-resolution zoom-in simulation with realistic stellar feedback processes. 
Although there could be other formation channels at high redshift \citep[e.g.,][]{kravtsov05,shapiro10,elmegreen12,kruijssen15},
we restrict our attention to the scenario based on small DMHs in this work.

\begin{figure*}
   \centering
         \includegraphics[width=18cm]{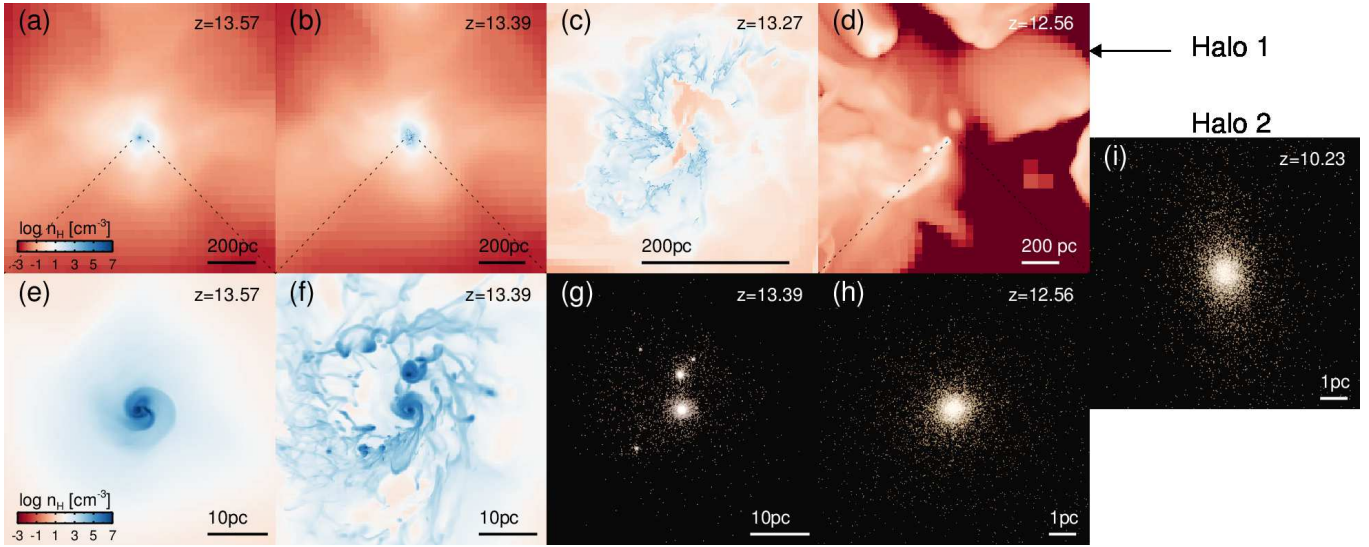}
   \caption{Different stages of GC formation in atomic-cooling halos. 
   Panels (a)-(f) and (g)-(h) show the gas distributions and 
   the NUV, B, V composite images of the stellar distributions generated using 
   {\sc sunrise} \citep{jonsson10} in Halo 1, respectively. 
   The stellar distribution for Halo 2 is shown in panel (i).
   When the mass of a halo is small enough that Ly$\alpha$ cooling is inefficient, 
   the halo gas cannot cool and accumulate at the center. Once the virial temperature 
   exceeds $\sim$ 8000 K, gas collapses isothermally, producing a gaseous disk (panels a and e). 
   Due to the short free-fall time of the star-forming gas, a large fraction of gas 
   is turned into stars before stellar feedback quenches star formation.
   During this stage, SNe exploding early ($\sim$ 3 Myr) enrich the surrounding medium,
   resulting in fragmentation of star-forming clouds (panels b and f). 
   We find that a significant fraction ($\sim 50\%$) of stars forms in these fragmented clouds (panel g).
   In $\sim$10 Myr, SNe effectively destroy star-forming clouds (panel c), 
   and prevent the halo from accreting gas for at least 100 Myr (panel d). 
   Star clusters produced in the fragmented clumps eventually merge into a single massive, 
   compact cluster (panel h).
   }
   \label{fig:pic}
\end{figure*}

%===================
\section{Numerical method}
%===================
We perform a cosmological radiation hydrodynamic simulation with the adaptive 
mesh refinement code, {\sc ramses-rt} \citep[][]{teyssier02,rosdahl15a}. 
The initial conditions are generated using {\sc music} software \citep{hahn11}, 
with the cosmological parameters 
($\Omega_{\rm m}=0.288$, 
$\Omega_{\rm \Lambda}=0.712$, 
$\Omega_{\rm b}=0.045$, 
$H_0=69.33\,{\rm km\,s^{-1}\,Mpc^{-1}}$, 
$n_s=0.971$, and 
$\sigma_8=0.830$) 
consistent with the WMAP9 results \citep{hinshaw13}.
The entire simulation box of (3 Mpc h$^{-1}$)$^3$ (comoving) is covered with 128$^3$ root cells.
High-resolution dark matter particles of mass 303\,\msun\ are adopted for 
a zoom-in region of 0.323 Mpc$^3$ (comoving), which encompasses two atomic-cooling halos 
with $4\times10^7\,\msun < M_{\rm halo} < 7 \times 10^7\,\msun$ at $z\sim10$. The zoom-in region is 
further refined to attain the maximum resolution of 0.1 pc (physical) 
if the total mass within a cell exceeds $2424\,\msun$ or if the gas mass in the cell with 
$\nH\ge 0.1\,{\rm cm^{-3}}$ exceeds $1.4\,\msun$. We also ensure that the shell formation 
radius of SN explosions is resolved by three cells \citep[e.g.,][]{kim15}, 
and that above densities of 10 H/cc, the Jeans length is resolved by 32 cells.
The Euler equations are solved using the second-order MUSCL scheme with an HLLC Riemann 
solver, and the Poisson equation is computed using a multi-grid method \citep{guillet11}.
For the transport of the three photon groups bracketed by ionization frequencies for {\sc HI}, {\sc HeI}, 
and {\sc HeII}, we adopt a GLF solver with a reduced speed of light approximation to reduce the computational costs \citep[$\hat{c}=10^{-3} c$,][where $c$ is the speed of light.]{rosdahl13}.

Star formation is modeled as a stochastic process \citep[][]{rasera06}, 
based on a Schmidt law \citep{schmidt59}, as
$\dot{\rho}_{\rm star} = \epsilon_{\rm SF} \rho_{\rm gas} / t_{\rm ff}$. Here, $\epsilon_{\rm SF}$ is 
the star formation efficiency per local free-fall time ($t_{\rm ff}=\sqrt{3\pi/32G\rho_{\rm gas}}$), 
which we take to be 2\% \citep{kennicutt98,krumholz07}.
The sites of star formation are limited to a dense ($\nH\ge10^5\,{\rm cm^{-3}}$) and converging gas 
flow ($\vec{\bf \nabla}\cdot \rho_{\rm gas} \vec{\bf v}_{\rm gas} < 0$), where $\vec{\bf v}_{\rm gas}$ is 
the gas velocity. The mass of each star particle (91 \msun) is chosen such that it hosts a single 
SN event for a Kroupa initial mass function (IMF) \citep{kroupa01}\footnote{This neglects the fact that 
one would require $\sim 10^3-10^{4}\msun$ to fully sample the IMF 
\citep[see][for a recent review]{kroupa13}. Choosing such a small mass for the star 
particle may lead to SNe exploding earlier than the realistic case and may over-predict the initial
chemical enrichment. However, given that a large amount ($10^6\,\msun$) of star-forming gas 
collapses on a very short timescale ($t_{\rm ff}\lesssim 0.1\,{\rm Myr}$) and 
stars amounting to $10^4\,\msun$ form quickly ($\sim 1\,{\rm Myr}$) in our simulations, 
we do not expect that fully sampling the IMF would significantly affect our main results on 
the stellar mass and metallicity distributions of the simulated clusters.}.
We assume that 21\% of the stellar mass is 
returned to the surroundings, among which 5\% in mass is the newly synthesized metals. 
This corresponds to a metal yield of $\approx$0.01 for an SSP with $1\,\msun$ .

We assume the initial metallicity of the simulation to be zero. 
Radiative gas cooling is computed by following the non-equilibrium chemistry of {\sc HI}, 
{\sc HII}, {\sc HeI}, {\sc HeII}, {\sc HeIII}, and $e^-$, coupled with the radiation. 
We also include metal cooling under the assumption of collisional equilibrium for $T\gtrsim10^4\,{\rm K}$  \citep{sutherland93}. Gas can cool down further with the metal fine-structure transitions \citep{rosen95}.  
Since our simulation explicitly follows the local 
ionizing radiation, we do not use the uniform ultraviolet background field.

Our simulation includes three different forms of stellar feedback.
First, ionizing photons from massive stars can 
heat up gas to $\approx 2\times10^4\,{\rm K}$ through photo-ionization, 
which subsequently reduces the gas density by over-pressurizing the surrounding medium. 
Second, absorption of ionizing photons from an SSP with $1\,\msun$ transfers 
the radial momentum of $\sim40\,\kms\,\msun$ to the surroundings. 
This is done by imparting momentum from ionizing radiation continuously 
for a given stellar metallicity and age, based on \citet{leitherer99}.
Finally, SN explosions are modeled by injecting the radial momentum 
calculated according to the stage of the Sedov-Taylor blast wave \citep{kimm14}. 
We also take into account the continuous spectrum of the lifetime of a massive star 
(from 3 Myr to 40 Myr) \citep{kimm15}. 
The simulations are run to $z=10.2$, by which two simulated GCs formed in atomic-cooling halos 
have become quiescent.

%===================
\section{Results}
%===================

In Figure~\ref{fig:pic}, we present the evolution of gas structures and the formation of two GC systems 
in atomic-cooling DMHs ($\mhalo\sim 4-7\times10^7\,\msun$) at high redshift ($z\gtrsim10$).
This may be summarized by five different evolutionary stages.
(1) When the halo mass is so small that its virial temperature is lower than $\sim 8000\,{\rm K}$, 
the gas in the mini-halo of primordial composition cannot cool through Ly $\alpha$ emission 
and, thus, remains rather diffuse ($\nH\lesssim100\,{\rm cm^{-3}}$) (adiabatic phase).
(2) Once a DMH becomes more massive than the transition mass 
$M_{\rm halo,tr} = 4\times10^7\,\left(\frac{1+z}{11} \right)^{-1.5}\,\msun$ through smooth accretion 
or halo mergers, the virial temperature of the halo exceeds $\sim 8000\,{\rm K}$ and the gas 
in the central region of the halo collapses nearly isothermally with the aid of strong 
Ly$\alpha$ cooling (panels a and e). Because the halo gas carries angular momentum 
generated by the large-scale tidal torque, the collapse leads to the formation of a dense, 
disc-like structure \citep[e.g.,][]{regan09} (isothermal cooling phase). 
(3) As the density of the disc becomes very high ($n_{\rm H} \gtrsim 10^6 \,{\rm cm^{-3}}$), 
its gas is turned efficiently into stars on a very short timescale ($\lesssim 10\,{\rm Myr}$). 
During this stage, early SNe ($t\sim3\,{\rm Myr}$) provide newly 
synthesized metals to the surrounding medium. The metals enhance radiative cooling in the 
star-forming disc, resulting in gas fragmentation (panels b and f).  
We find that roughly half of the stars (56 and 50\%, respectively) in the simulated GCs form 
in the fragmented clumps (panel g) (self-enrichment and fragmentation phase). 
(4) A large number of SN explosions ($\sim10^4$) ensue, clearing out the star-forming 
clouds and gas in the central region of the halo (panel c) (outflow phase).
(5) Subsequent star formation is quenched for at least one hundred Myr, and star clusters at the center of 
the halo quickly merge into a single cluster. Then, dynamical interactions of stars inside each cluster are 
likely to govern the structural evolution from this point \citep[e.g.,][]{gao91}.

\begin{figure}
   \centering
   \includegraphics[width=8.5cm]{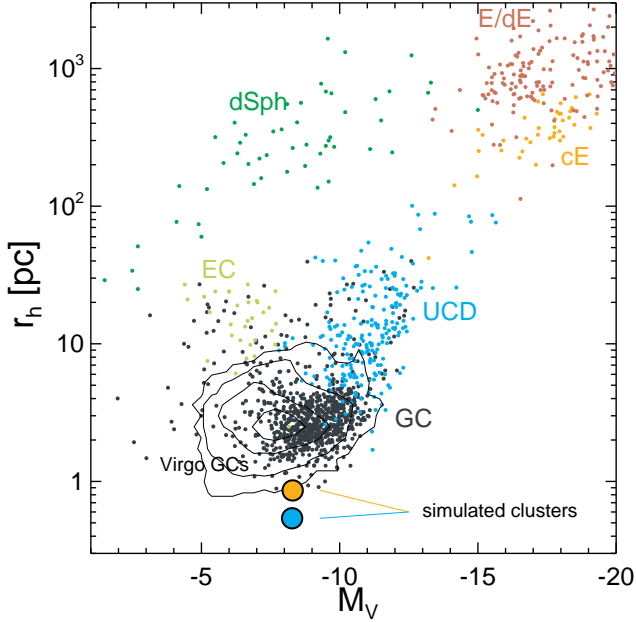}
   \caption{Comparison of half-light radii and $V$-band magnitudes between the two simulated GCs 
   (filled circles with the black envelope) and the observations compiled by \citet{brodie11}. 
   Data points are color-coded according to their structural types (GC: globular clusters, UCD: ultracompact dwarves, EC: extended clusters, dSph: dwarf spheroidal, 
   dE: dwarf ellipticals, cE: compact ellipticals). 
   Also included as black contours is the distribution (10, 68, 95, 99.5\%) of the GCs from the ACS 
   Virgo Cluster Survey \citep{jordan09}. We use the empirical color transformation of the GCs to convert 
   from $g$ and $z$ bands to $V$ band \citep{forbes13}. The size of the simulated clusters is measured  
  when a single massive cluster is formed after mergers of clusters in the disk. 
  The V-band magnitudes of the simulated GCs are 
   calculated assuming that member stars evolve to $z=0$ without any loss due to dynamical interactions. }
   \label{fig:size}
\end{figure}

\begin{figure}
   \centering
   \includegraphics[width=8.5cm]{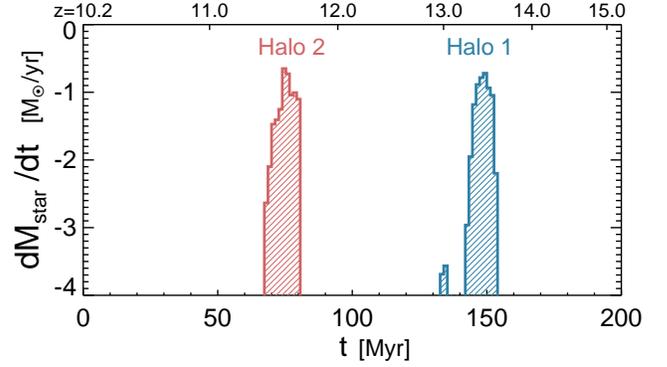}
   \caption{Star formation histories of the two simulated GCs. Stars form very efficiently 
   within $\lesssim$ 10 Myr before SNe effectively destroy star-forming clouds and suppress gas accretion. 
   The resulting star clusters remain as an SSP within the timescale of 
   our simulation.
     }
   \label{fig:sfh}
\end{figure}

We find that the resulting dense clusters in the two halos share two important 
properties with the local GC populations.  First, the mass of the simulated GCs 
($6\times10^5\,\msun$ for both halos) is comparable to the local GCs \citep{harris96,strader11}. 
If we assume that the stars in these systems are unaffected 
by tidal stripping and simply evolve to $z=0$, their $V$-band magnitudes would 
also be consistent with observations (Figure~\ref{fig:size}). Here we compute the 
$V$ band flux assuming a Kroupa IMF with a low- (high-) mass cut-off of 0.1 (100) 
\msun\ \citep{leitherer99}. Second, the simulated GCs are found to be very compact.
When mergers of several clusters are completed and form a single massive cluster, 
the three-dimensional half-light radius is 0.54 and 0.86 pc, respectively. 
Although the size of the simulated clusters is approximately a factor of 3--5 smaller than 
the observational estimates \citep[e.g.,][]{harris96,jordan09,brodie11},
we note that subsequent dynamical evolutions through binary interactions 
\citep[e.g.,][]{gao91} or mass loss are likely to increase the size of the systems.
In a similar context, \citet{baumgardt10} suggests that the initial size of the local 
compact GCs is likely to be less than 1pc \citep[c.f.][]{shin13}\footnote{Note that the accurate determination of the initial size of GCs will 
require the inclusion tidal shocks due to giant molecular clouds, which is not taken 
into consideration in \citet{baumgardt10}.}.
It is also worth mentioning that each simulated cluster is formed in a single burst of 
star formation ($\lesssim 10\,{\rm Myr}$) (Figure~\ref{fig:sfh}), 
roughly consistent with the fact that a GC is essentially an SSP\footnote{We note, however, 
that the simulated clusters would have formed on a timescale shorter than $\sim$ 10 Myr 
if the effect of radiation feedback was stronger. 
Due to the lack of cooling agents in our primordial halos,
the temperature of the star-forming gas does not drop far below $\sim10^4$ K,
and the external pressure around the star-forming cloud is over-estimated.
As a result, HII bubbles are confined by the surroundings and cannot drive outflows in our simulation.
Including relevant cooling processes by molecular hydrogen and metals produced by Pop III/II stars will 
decrease the temperature of the star-forming regions and 
result in an enhanced photoionization feedback  before SNe explode.
This is likely to reduce the star formation timescale, 
which will be more compatible with the age spreads found in the local young massive clusters ($\lesssim$ 3 Myr).
}.
Note that such a sharp truncation in star formation is driven by efficient SN feedback.

\begin{figure}
   \centering
      \includegraphics[width=8.5cm]{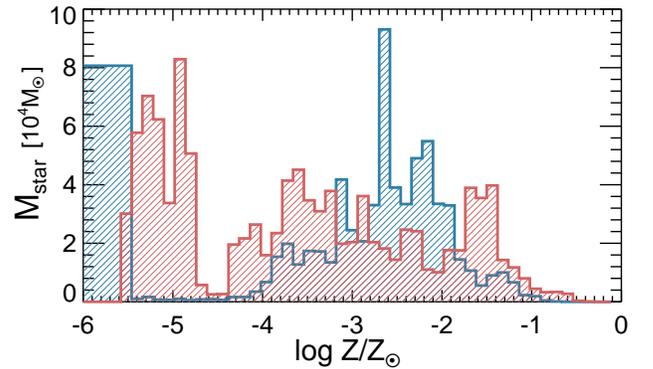}
   \caption{Metallicity distributions of stars in the two simulated GCs. 
   Different colors denote different GCs, as in Figure~\ref{fig:sfh}.
   The mean metallicities of the stellar systems are  0.006 and 0.012 \zsun, respectively.
   Metallicities lower than $10^{-6}\,\zsun$ are shifted to $10^{-6}\,\zsun$ in Halo 1.
   Note that there are no stars with zero metallicity in Halo 2, 
   as they are pre-enriched by the SN ejecta from Halo 1. 
   Although a fraction of stars is quickly enriched to 0.1 \zsun,
   we find that there is a wide spread in metallicities in contrast to observations.
   }
   \label{fig:met}
\end{figure}

\begin{figure}
   \centering
         \includegraphics[width=8cm]{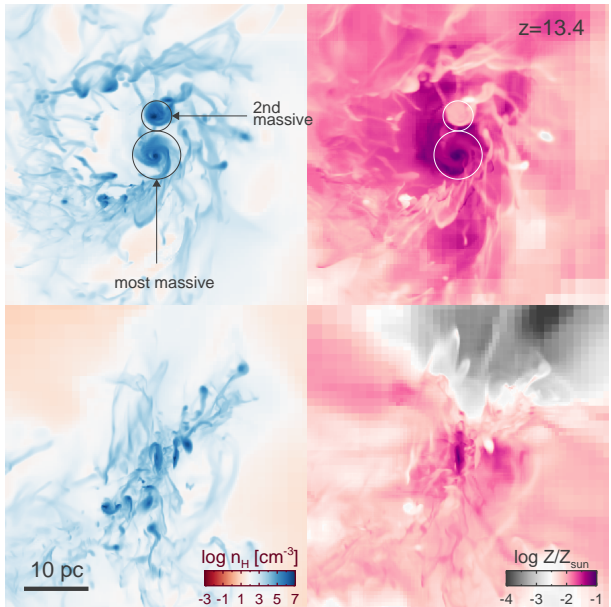}
   \caption{Inhomogeneous distributions of metals during the self-enrichment by SN explosions. 
   The mass-weighted density and metallicity distributions are shown in left and right panels, 
   respectively. Top and bottom panels are different projections (x, z). 
   Note that dense star-forming clumps ($\nH\ge 10^5\,{\rm cm^{-3}}$) exhibit a wide range of metallicities.
   The non-uniform metal enrichment appears inevitable in the scenario
   where GCs form from pristine halo gas, posing a challenge to GC formation theory.  }
   \label{fig:metfig}
\end{figure}

Although the predicted size and mass are comparable with their local counterparts, 
we find that the simulated GCs show a wide spread in metallicities.
The mass-weighted mean metallicity of the two GC candidates (0.006 and 0.012 $\zsun$) is 
close to that of the local metal-poor populations, but the metallicity of individual 
star particles is found to range from zero to a tenth of Solar (Figure~\ref{fig:met}).
Such a large dispersion is not unexpected because our simulations start from zero metallicity.
Yet it is interesting to note that the fraction of the extremely metal-poor population 
($Z \lesssim 10^{-3}\,\zsun$) is significant (50--60\%) in contrast to the standard interpretation 
that a GC is rather a chemically homogeneous system \citep{harris96}. 
This can be attributed to the fact that, in simulations, a large fraction of gas is converted into 
stars before the first SN emerges and pollutes the star-forming gas. Moreover, as metals from 
SNe are dispersed locally and some of SNe exploding on the outskirt of star-forming clouds can simply 
escape from them, we find that the metal enrichment occurs in a very inhomogeneous fashion.
For example, Figure~\ref{fig:metfig} shows that the second massive star-forming cloud 
(upper giant cloud in the first panel) is at least an order of magnitude more metal-poor 
than the central gas clump.
However, it should be noted that the metallicity distributions may not necessarily rule out 
the DMH-based scenario, because pre-enrichment by Pop III and II 
stars can reduce the metallicity spread by effectively regulating the self-enrichment via
radiation feedback. It is also possible that the inclusion of turbulent metal mixing \citep[e.g.,][]{shen10} 
that is not fully resolved in this study would suppress the formation of high- and low-metallicity stars. 
We discuss a role of pre-enrichment by the first generation stars and metal mixing
as a potential solution to this problem in the next section.

\section{Discussion and Conclusions}

Using cosmological radiation hydrodynamic simulations, we show that a massive and compact 
star cluster can form at the center of a DMH during the transition from molecular-cooling 
to atomic-cooling halos, provided that gas is primordial and atomic.
Because gas collapse occurs isothermally under such conditions, a large fraction of 
the gaseous disc with very high density turns into stars before SNe disrupt the star-forming clouds. 
During this stage, the gaseous disc is fragmented and stars form in each clump
due to self-enrichment from early SNe commencing at 3 Myr, unlike the picture of black 
hole seed formation in which gas simply collapses onto AU scales \citep[e.g.,][]{choi13}. 
The resulting star clusters are found to have similar physical properties (size and mass) 
to the local GC populations, except for metallicity spread and the possibly related star-formation timescale.  

The most critical assumption in this scenario is that gas is primordial 
and atomic so that cooling due to molecular hydrogen and subsequent star formation 
does not prevent a large amount of gas from accumulating in the central region of the halo to form a GC.
This may be achieved by having a strong Lyman-Werner background radiation that photo-dissociates 
molecular hydrogen \citep[e.g.,][]{omukai01}. However, we argue that even with cooling due to 
molecular hydrogen or metals produced by neighboring galaxies, it may still be possible to 
form a dense cluster, provided that Pop III or Pop II populations effectively regulate gas accretion. 
If gas is allowed to collapse in mini-halos via molecular hydrogen cooling 
and form Pop III stars \citep{greif10}, 
SN explosions and radiation feedback may expel a large fraction of the halo gas, 
suppressing star formation.
If stellar feedback is energetic enough, subsequent major gas collapse is likely to occur
when a large amount of gas is available and radiative cooling becomes significant, 
i.e., when the halo mass becomes close to the atomic-cooling regime. 
Even in the case where the recovery timescale of gas accretion is short due to 
weak feedback from Pop III stars, the first generation of Pop II stars can in principle blow 
away the interstellar medium without constituting the GC population 
if the ensuing collapse of gas does not occur exactly inside the first Pop II stars (Kimm et al. 2016, in prep.).
Given that Pop III stars can pre-enrich the halo gas to $Z\sim10^{-3}-10^{-2}\,\zsun$ \citep{boley09,greif10,ritter12} and  halos with a few times $10^7\,\msun$  are 
likely to produce $\sim100\,\msun$ of metals \citep[e.g.,][]{wise14}, 
it can also help to alleviate the metallicity spread found in our simulation,
possibly explaining the helium enhancement observed in some GCs \citep[][]{choi07}.

We note that the DMH-based scenario may be able to accommodate the observed 
number of blue GCs if a significant fraction of DMHs forms a GC during the transition 
from molecular-cooling to atomic-cooling regime. To estimate the {\it maximum} frequency, 
we generate halo merger trees for the volume of $(100\,{\rm Mpc\,h^{-1}})^3$ using the 
Monte Carlo algorithm based on the extended Press Schechter theory \citep{parkinson08}. 
We adopt the same cosmological parameters as described in Section~2 and 
a mass resolution of $M_{\rm res}=5\times10^6\,\msun$. Then the number of halos 
that first cross the transition mass ($\mhalo = 4\times10^7 \left(\frac{1+z}{11}\right)^{-1.5} \,\msun$) 
is computed throughout the merger history in each halo. We find that the resulting frequency is
$N_{\rm GC,max}\approx650\pm20\times \left(\mhalo / 10^{12}\,\msun\right)$.
Thus, for a Milky Way hosting halo of mass $10^{12}\,\msun$, the DMH-based formation scenario 
can predict the number of GCs up to six times more than the number of the local 
metal-poor population \citep[e.g.,][]{harris96} or an order of magnitude larger stellar masses 
associated with GC systems per dark matter halo mass \citep{spitler09,hudson14,durrell14}. 
The large number of GCs may seem inconsistent with the local values, 
but it should be noted that the actual abundance at $z=0$ is likely to be lower than the maximum 
estimate for several reasons.
It is known that the GC system can be reduced or disrupted by two-body relaxation-driven, 
tidally limited evaporation \citep{fall01,vesperini01}, disc and bulge shocks \citep{gnedin97}, 
and tidal shocks by giant molecular clouds in the disc \citep[][]{kruijssen12,kruijssen15}.
It is also possible that the system ends up 
as a nuclear star cluster instead of a GC, given that it is embedded in the central region of a DMH.
More importantly, not all of the dark matter halos may host a GC if gas accretion 
takes place rather continuously due to the inefficient explosion of the first generation stars 
and form a dwarf galaxy. 
In this regard, we argue that the large maximum frequency of GCs simply indicates that
the DMH-based scenario can easily accommodate the abundance of local blue GCs.

Finally, it should be mentioned that the DMH-based scenario may be 
ruled out if overall metal enrichment from the first generation stars (i.e. Pop III and Pop II) 
or external stellar sources \citep[][]{trenti15} is not sufficient enough to reach the typical 
metallicity of metal-poor GCs ($Z\sim0.02\,\zsun$). As shown in Figures 4 and 5, 
the self-enrichment by stars in the simulated GCs is highly inhomogeneous 
and leads to a wide stellar metallicity distribution of $10^{-5}\lesssim Z/\zsun \lesssim 10^{-1}$, 
in contradiction to the observed metallicity spread \citep[][]{gratton12}. 
Part of the large spread may be caused by the neglect of turbulent metal mixing that 
is not fully captured in our simulation due to limited resolution. 
The inclusion of explicit turbulent metal mixing \citep[e.g.,][]{shen10} is likely to
spread out the metal from highly enriched regions ($Z\gg 0.02\,Z_{\odot}$) 
and increase the metallicity in metal-poor regions. However, the mixing alone may 
not be able to solve the metallicity problem, given that low-metallicity stars would 
still form before the first SNe explode and enrich the interstellar medium.
Instead, we note that pre-enrichment by Pop III and II stars can significantly reduce 
the metallicity spread by removing the low-metallicity tail. 
This pre-enrichment, along with the cooling by molecular hydrogen, can also effectively 
suppress the formation of high-metallicity stars for the following reason.  
Due to the lack of cooling agents in our primordial halos, we find that the temperature of star-forming gas 
does not drop far below 8000K during the early stage of star formation, making the external 
gas pressure in the star-forming regions too high. 
This leads to an underestimation of the effect of photoionization feedback,
because the HII region is no longer an over-pressurized bubble and cannot generate
enough momentum to disperse gas clouds on a very short time scale of a few Myr \citep{walch12,dale14}.
However, including the relevant cooling processes by metal and molecular hydrogen 
will lower the temperature of the star-forming region, and the inner pressure from HII regions 
will be able to drive outflows and suppress star formation 
before newly synthesized metals from Type II SNe ejecta is recycled to star-forming clumps.
Thus, the metallicity problem may be solved 
by taking into account Pop III stars and relevant cooling processes.
In this case, the self-enrichment process will be limited accordingly, 
and one would require the pre-enrichment of metals that closely 
match those of local blue GCs.
If a wide range of pre-enrichment by Pop III stars is real 
\citep[$Z\sim10^{-3}-10^{-2}\,\zsun$,][]{boley09,greif10,ritter12} and 
 only a fraction of the metal ejecta from the first Pop II stars is 
recycled due to outflows, an important question 
would then be why GC formation is efficient only in regions with $Z\gtrsim0.02\,\zsun$.
Future studies with Pop III stars and full-blown chemistry will be useful to 
more accurately assess the formation scenario of GCs at the center of DMHs.

\acknowledgements{
We are indebted to an anonymous referee for a constructive and thorough report 
that improved the presentation of our work.
We thank Kengo Tomida for useful discussions, and Romain Teyssier 
for making his code {\sc ramses} publicly available. 
Computing resources were in part provided by the NASA High-End 
Computing (HEC) Program through the NASA Advanced
Supercomputing (NAS) Division at Ames Research Center and in 
part by the Horizon-UK program through DiRAC-2 facilities.
The research is supported in part by NSF grant AST-1108700 and NASA grant NNX12AF91G
and in part by the ERC Advanced Grant 320596 `The Emergence of Structure during the 
epoch of Reionization'. 
JR was funded by the European Research Council under the European Unions Seventh Framework Programme (FP7/2007-2013)/ERC Grant agreement 278594-GasAroundGalaxies, and the Marie Curie Training Network CosmoComp (PITN-GA-2009-238356). SKY acknowledges support from the 
Korean National Research Foundation (Doyak 2014003730).
}

\small
\bibliographystyle{apj}
%\bibliography{refs}

\end{document}